\documentclass[sigconf, screen, natbib=false, nonacm]{acmart}
\AtBeginDocument{%
  
}

\usepackage{tabularx}
\usepackage{xurl}
\usepackage{hyperref}
\usepackage{xcolor}
\usepackage{enumitem}

\newcommand{\systemname}{\texttt{Eliot}}

\usepackage{xcolor}

\usepackage{xcolor}

\RequirePackage[
  datamodel=acmdatamodel,
  style=acmnumeric,
  ]{biblatex}

\begin{document}

\title{Eliot: Interactively \underline{E}xploring Fast-Changing Scientific \underline{Li}terature Trends with \underline{O}nline Da\underline{t}a and Learning}

\author{Bernardo A. Denkvitts}
\affiliation{%
  \institution{University of South Carolina}
  \city{Columbia}
  \state{South Carolina}
  \country{USA}
}
\email{denkvitb@email.sc.edu}

\author{Nitin Gupta}
\affiliation{%
  \institution{University of South Carolina}
  \city{Columbia}
  \state{South Carolina}
  \country{USA}
}
\email{niting@email.sc.edu}

\author{Biplav Srivastava}
\affiliation{%
  \institution{University of South Carolina}
  \city{Columbia}
  \state{South Carolina}
  \country{USA}
}
\email{biplav.s@sc.edu}



\begin{abstract}
  The rapid growth of scientific publishing has made it increasingly difficult to track how fast-moving areas evolve.
  Search engines and LLM-based assistants retrieve or summarize papers, but often hide how the corpus was selected, organized, or connected to temporal patterns.
  We present \systemname, a publicly deployed interactive system for traceable exploration of evolving scientific literature.
  Motivated by two studies on Large Language Models (LLMs) and Automated Planning and Scheduling (APS), \systemname{} generalizes literature-evolution analysis beyond hand-built taxonomies and domain-specific scripts.
  Given explicit query terms and filters, it retrieves arXiv papers at query time, represents each paper by title and abstract, clusters the corpus into themes, assigns representative keywords, and visualizes each cluster's publication-year distribution.
  We evaluate \systemname{} as both an applied system and an interactive research aid.
  An offline configuration study across eight arXiv domains compares document representations, dimensionality reduction methods, and clustering algorithms using intrinsic clustering and topic-coherence metrics; the results support MiniLM embeddings with 10-dimensional UMAP and Agglomerative Clustering as a practical default.
  A scenario-based survey and expert focus group assess interpretability and use contexts: participants rated cluster labels as meaningful in 85\% of scenario responses, and feedback indicated that \systemname{} is most valuable for auditable overviews of rapidly changing technical areas.
  These results suggest that query-time clustering and temporal inspection can complement search and generation tools by helping researchers inspect and refine the evidence behind literature trends.
\end{abstract}



\raggedbottom


\maketitle

\section{Introduction}
\label{sec:introduction}

\begin{figure*}
    \centering
    \includegraphics[width=.8\linewidth]{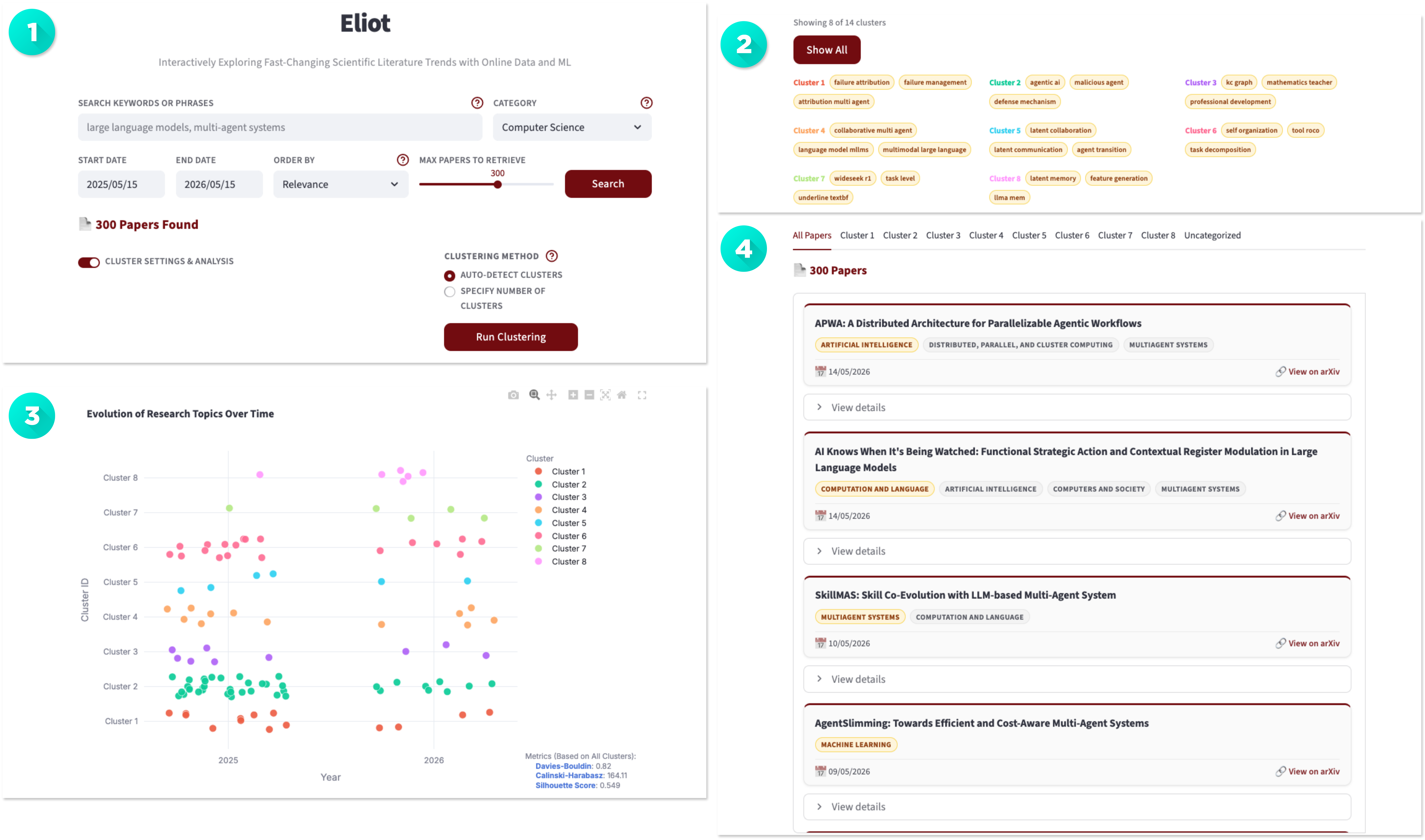}
    \caption{The \systemname{} user interface, shown across four screenshots of a single scrollable application. (1) The search panel with default run configurations. (2) Auto-detected cluster overview, showing representative keywords for 8 of 14 discovered clusters (the full set is accessible via `Show All'). (3) Temporal interactive scatter plot showing the evolution of research topics over time across clusters. (4) Paginated paper list with per-cluster tabs. Additional papers are not shown due to space.}
    \label{fig:app}
\end{figure*}

Scientific publishing is growing quickly, making it difficult to track fast-changing areas~\cite{probierz_clustering_2022,hanson_strain_2024}. Researchers must find relevant papers, understand how they separate into themes, identify emerging directions, and verify which documents support an apparent temporal pattern~\cite{pallagani2025revisiting,zhang_survey_2018}. Recent literature assistants make search more interactive~\cite{schneider_conversational_2024,bian2024intellectseeker}, but ranked lists and generated syntheses can hide retrieval, selection, and organization decisions. Literature evolution analysis therefore needs an inspectable workflow in which users can see the corpus, themes, labels, temporal patterns, and source papers.

We present \systemname, an interactive system for exploring evolving scientific literature from online data.\footnote{Here, \emph{online data} means that the corpus is retrieved at query time from arXiv rather than drawn from a fixed benchmark dataset, and \emph{trends} refers to inspectable temporal patterns in that retrieved corpus rather than fully automated trend discovery.} Given keywords, category filters, date ranges, and sorting criteria, \systemname{} retrieves papers from arXiv at query time, clusters them, labels clusters with representative keywords, and visualizes temporal distributions. The public Streamlit prototype\footnote{\url{https://ai4society-paper-searcher.streamlit.app}} emphasizes traceability: users can revise the query, inspect cluster memberships, compare labels, and open source records rather than relying only on an opaque summary.

\systemname{} is motivated by two ICAPS studies on LLMs in Automated Planning and Scheduling (APS), or \emph{LLM-planning}. \textcite{pallagani2024prospects} manually reviewed 126 papers into eight categories, including plan generation, language translation, model construction, tool integration, and interactive planning. \textcite{pallagani2025revisiting} then added a semi-automated arXiv pipeline for fetching, deduplication, classification, and trend analysis. With 47 new papers, it reported category drift: six categories declined in relative share, two increased, and goal decomposition and replanning emerged. These studies showed that category-level literature evolution is useful, but the workflow required a fixed taxonomy, domain-specific scripts, supervised labels, expert review, and post-hoc temporal analysis.

\systemname{} generalizes this workflow from one expert-maintained domain to arbitrary arXiv-supported topics. Rather than classifying papers into prebuilt categories, it performs unsupervised, query-time grouping over the retrieved corpus, labels the discovered clusters, and exposes the source evidence behind each trend. This shift motivates our evaluation of semantic representations and clustering configurations beyond the TF-IDF supervised setting used in the \emph{LLM-planning} update.

Our evaluation is organized around three empirical questions: \textbf{(RQ1)} \textit{Which automated configuration should \systemname{} use by default?} We ask which combination of document representation, dimensionality reduction, and clustering provides robust structure across diverse arXiv domains. \textbf{(RQ2)} \textit{Do users find the generated clusters interpretable and useful?} We ask whether users perceive the cluster labels, groupings, and interaction flow as meaningful for literature exploration. \textbf{(RQ3)} \textit{When do users expect \systemname{} to be most useful, and what limitations matter?} We ask which research contexts benefit from traceable cluster-and-time views, and where users need additional guidance when interpreting changing corpora.
Thus, our contributions are:
\begin{itemize}[noitemsep, leftmargin=1.5em]
    \item A public interactive system for literature evolution exploration that combines query-time arXiv retrieval, clustering, keyword labeling, temporal visualization, and document-level inspection.
    \item A traceable workflow that exposes query settings, retrieved papers, cluster assignments, and source links, supporting auditability in exploratory literature analysis.
    \item An automated offline configuration evaluation across eight arXiv domains, complemented by a scenario-based user survey and expert focus group.
\end{itemize}

\section{Related Work}
\label{sec:related_work}

\begin{figure*}[t]
    \centering
    \includegraphics[width=.9\textwidth]{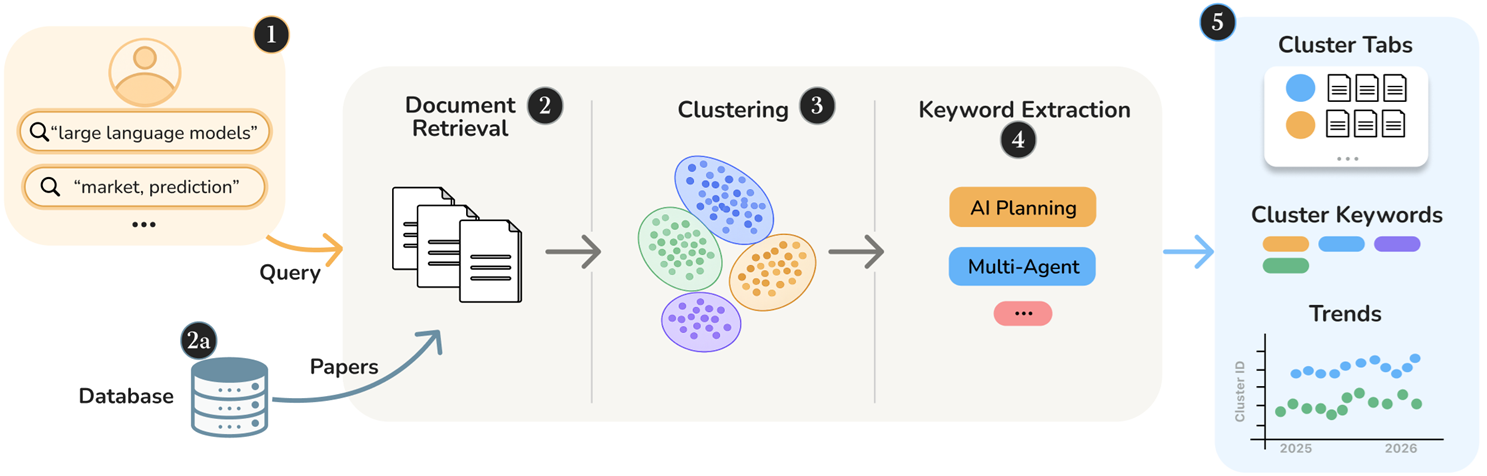}
    \caption{System pipeline overview. (1) The tool supports multiple keywords, both single-word and multi-word expressions. (2, 2a) After receiving user keywords, the system retrieves papers from arXiv. (3) It performs clustering and (4) keyword extraction. (5) The UI displays cluster keywords, cluster tabs, and a temporal visualization to support structured exploration.}
    \label{fig:system_pipeline}
\end{figure*}

This section situates \systemname{} within four lines of work: \emph{LLM-planning} literature analysis, scientific text clustering, AI-assisted search and synthesis, and literature discovery systems.

\subsection{LLMs in Planning as a Motivating Case}
The \emph{LLM-planning} literature illustrates why fast-changing fields need traceable evolution tools. \textcite{pallagani2024prospects} manually reviewed 126 papers and proposed an eight-category taxonomy covering language translation, plan generation, model construction, multi-agent planning, interactive planning, heuristics optimization, tool integration, and brain-inspired planning. The survey connected category-level volume to challenges such as executable plan generation, grounding language in planning representations, coordinating agents, and integrating LLMs with verifiers, planners, or other tools.

\textcite{pallagani2025revisiting} revisited the area with a semi-automated arXiv pipeline for extraction, deduplication, categorization, and supervised classification experiments. The study added 47 papers, reported drift across the original taxonomy, and surfaced goal decomposition and replanning as emerging categories. Its strongest result was human-augmented, showing that automation can scale updates but still needs inspectable evidence and expert judgment.

\systemname{} starts from this lesson but changes the task. The ICAPS workflow assumed a known domain, fixed taxonomy, labeled examples, and bespoke scripts. \systemname{} supports earlier-stage exploration: it retrieves a fresh corpus from explicit query settings, clusters documents without supervised labels, produces representative keywords, and connects temporal patterns back to individual papers. The \emph{LLM-planning} case is therefore a motivating precursor rather than a directly comparable baseline.

\subsection{Scientific Text Clustering}
The rapid growth of scientific publications has motivated unsupervised methods for organizing academic texts.
\textcite{probierz_clustering_2022} compare TF, TF-IDF~\cite{qaiser_text_2018}, and binary weighting with K-Means~\cite{mcqueen1967some}, finding that TF-IDF performs best under their keyword-overlap connection coefficient and that introductions can yield stronger clusters than abstracts. \systemname{} uses titles and abstracts because they are concise and computationally lighter for interaction. Since the \emph{LLM-planning} update used TF-IDF for supervised classification into a fixed taxonomy~\cite{pallagani2025revisiting,pallagani2024prospects}, we ask whether that choice generalizes to unsupervised, domain-agnostic exploration.

\textcite{cherradi_exploration_2024} compare K-Means, Hierarchical Agglomerative, and DBSCAN using intrinsic metrics such as Silhouette Score (SIL), Calinski-Harabasz Index (CHI), and Davies-Bouldin Index, reporting DBSCAN as strongest on SIL and K-Means on CHI. Transformer-based work similarly seeks better semantic representations: \textcite{taskiran_academic_2022} find that BERT~\cite{devlin_bert_2019} outperforms Word2Vec for Turkish academic abstracts, while \textcite{weng_identification_2022} use GPT-3 similarity embeddings with HDBSCAN for scientific abstracts.

Despite these advances, existing studies present two limitations for our applied setting. First, they primarily rely on intrinsic clustering metrics that do not capture whether cluster labels are interpretable to users. Second, they are typically evaluated on fixed corpora, without assessing whether a configuration is robust across domains likely to be explored by a public tool.
To address these gaps, we perform a systematic evaluation of text representation methods, dimensionality reduction techniques, and clustering algorithms. Our evaluation combines intrinsic clustering metrics with topic coherence measures ($C_V$ and $C_{\text{NPMI}}$), and analyzes the stability of configurations across eight scientific domains.

\subsection{AI-Assisted Literature Search and Synthesis}
Recent AI research assistants make scholarly search more interactive by accepting natural-language questions, expanding search strategies, and synthesizing evidence with citations. \textit{Scholar Labs} analyzes detailed research questions into topics, aspects, and relationships, searches Google Scholar for those facets, and provides per-paper explanations~\cite{google_scholar_labs_2025}. \textit{Undermind} similarly frames literature search as a multi-step investigation: its white paper describes a language-model-guided pipeline that combines semantic search, citation trails, relevance classification, adaptive exploration, and estimates of search completeness~\cite{hartke2024benchmarking}. While its white paper describes the initial pipeline over arXiv full texts, it has since expanded to a broader Semantic Scholar corpus. Other tools emphasize evidence synthesis and verification. \textit{Elicit} supports systematic-review stages such as protocol refinement, paper gathering, screening, extraction, and report generation~\cite{elicit_systematic_reviews_2026}. \textit{Consensus} combines semantic and keyword retrieval over a large literature database with AI summaries, quality signals, and agreement visualizations~\cite{consensus_how_2026}. \textit{Scite} complements search by classifying citation contexts as supporting, contrasting, or mentioning a cited claim~\cite{nicholson2021scite}.

Research prototypes push this direction further with retrieval-augmented agents. \textit{PaperQA} and \textit{PaperQA2} retrieve full-text sources, assess passages, and generate cited answers for scientific questions~\cite{lala2023paperqa,skarlinski2024language}. \textit{OpenScholar} answers scientific queries by retrieving passages from 45 million open-access papers and producing citation-backed long-form responses~\cite{asai2024openscholar}. These systems are important comparators because they address the same high-level problem: researchers need help searching, reading, and synthesizing rapidly growing literature.

Our goal is narrower and more inspectable. \systemname{} does not attempt to produce a final evidence synthesis, screen papers for a formal systematic review, or decide which claim is true. Instead, it supports an earlier exploratory task: making the retrieved corpus visible as themes, labels, temporal patterns, and paper-level evidence. This distinction matters because answer-oriented assistants can compress retrieval, selection, ranking, and synthesis into a generated response. \systemname{} exposes these intermediate objects directly, thereby complementing AI assistants that could later improve query expansion, paper summarization, or claim verification.

\subsection{Literature Discovery and Exploration}
The growing volume of scientific publications has also motivated discovery systems. \textcite{schneider_conversational_2024} propose a conversational search system built on an ACL Anthology knowledge graph with precomputed clusters and keyword labels; because its analysis is computed offline over a fixed corpus, it cannot dynamically adapt to emerging topics. \textcite{katz2024knowledge} present \textit{Knowledge Navigator}, which organizes documents from broad scientific queries into a two-level hierarchy of named topics and subtopics. This is close to our emphasis on structured overviews, but \systemname{} instead provides a deployed, query-time workflow linking clusters to keywords, temporal distributions, and inspectable paper records.

\textcite{bian2024intellectseeker} introduce \textit{IntellectSeeker}, a personalized literature management platform that combines probabilistic recommendation with LLM-based semantic enhancement and dynamic web acquisition. However, beyond summaries and word clouds, it provides limited structure for analyzing retrieved documents. Citation-network tools such as \textit{Litmaps} and \textit{ResearchRabbit} help users expand from seed papers and visualize connections~\cite{litmaps_researchers_2026,researchrabbit_2026}. \systemname{} instead starts from explicit query terms and arXiv filters, then constructs semantic clusters and temporal views over the retrieved set.

Prior work on research trend analysis often studies topic evolution and emerging directions using static datasets and offline analysis~\cite{chen_citespace_2006, wang_itginsightdiscovering_2022, pallagani2025revisiting}. These systems demonstrate the value of temporal views, but they usually require precomputed corpora or expert-created taxonomies. \systemname{} instead integrates dynamic retrieval, on-the-fly clustering, keyword labeling, temporal visualization, and deployment into a traceable workflow for researchers working with changing corpora.

\section{System Overview}
\label{sec:system_overview}

\systemname{} is designed for exploratory analysis of scientific literature. It combines query-time document retrieval, machine-learning-based clustering, keyword labeling, temporal visualization, and document inspection so that users can move from a broad query to an evidence-backed view of themes in the retrieved corpus. An overview of the pipeline is shown in Figure~\ref{fig:system_pipeline}.
This section outlines overall system design and implementation details for \systemname{} are provided in Section~\ref{sec:implementation}.


\systemname{} is guided by \textbf{four design goals}. First, it should support \emph{query-time exploration} so users can study fast-changing topics without waiting for a curated corpus or maintaining a domain-specific extraction script. Second, it should provide \emph{structured overviews} by grouping retrieved papers into themes rather than showing only a ranked list. Third, it should support \emph{temporal inspection} by showing how clustered papers are distributed across publication years, while leaving trend interpretation auditable by the user. Fourth, it should preserve \emph{traceability}: every theme and temporal pattern should remain connected to the papers, abstracts, metadata, and source links that produced it.

The \textbf{system workflow} retrieves a set of relevant documents from the paper database given a query composed of keywords or comma-separated phrases, along with filters such as paper category, date range, and sorting criterion. The query is intentionally explicit: each term is searched in titles or abstracts, category filters are visible, and the date range defines the temporal window used in downstream analysis. Keeping retrieval terms explicit supports traceability and makes the system complementary to automatic query expansion methods.

The retrieved documents are represented using their titles and abstracts, then grouped into clusters of semantically related papers. This changes the workflow from the \emph{LLM-planning} studies: users no longer need to define 8, 10, or 12 categories before analysis begins. Instead, they can let the system infer a structure from the current result set or specify a cluster count when they want to compare a coarser or finer thematic view. To enhance interpretability, each cluster is associated with representative keywords that summarize what distinguishes it from the rest of the retrieved corpus. The keyword layer is important because users first encounter clusters through short labels, then decide which groups warrant deeper inspection.
Users can influence the clustering process through two modes: \textit{automatic clustering}, where the system estimates clusters from the structure of the retrieved articles, and \textit{user-controlled clustering}, where users specify the desired number of clusters, supporting sensitivity analysis when the automatic grouping is too coarse or too fragmented.

The interface presents the result count, cluster labels, temporal scatterplot, quality metrics, and cluster-specific tabs. Within each tab, users can inspect titles, metadata, abstracts, and links to the original database records. This design keeps the system grounded in documents rather than generated summaries, which is essential for trust in exploratory literature analysis.

\section{Implementation Details}
\label{sec:implementation}

This section describes the deployed prototype and the offline evaluation datasets. The user interface is shown in Figure~\ref{fig:app}. A central implementation goal is traceability: the interface exposes query settings, retrieved papers, cluster assignments, representative keywords, metrics, and temporal visualizations rather than only returning a generated narrative. The machine-learning layer is unsupervised and is used to structure query results through text representation, dimensionality reduction, clustering, and cluster-keyword extraction.

ArXiv was chosen as the data source due to its open access nature, broad coverage of scientific topics, and up-to-date content. The system constructs queries, maps arXiv subject categories, and retrieves papers using Python packages~\cite{lukas_arxiv_api, roman_arxivql_api}. User keywords are interpreted as comma-separated terms; each term is searched in the title or abstract, and all terms are combined conjunctively. Multi-word expressions are preserved as phrases. Category and date filters are applied. The API returns the latest version of each paper.
The deployed search client further allows the option to choose the maximum number of results (20--500) from arXiv, while the offline evaluation datasets in Table~\ref{tab:datasets} are capped at 300 papers per domain to keep configuration sweeps comparable and computationally manageable.

For text processing, the title and abstract are concatenated for each paper. The deployed prototype uses a lightweight text preprocessing pipeline: text is lowercased, lightly cleaned, part-of-speech tagged, and lemmatized. Then, for text representation, Sentence-Transformers all-miniLM-L6-v2 \cite{reimers_sentence-bert_2019} is applied to the preprocessed texts before clustering.


Clustering in the deployed prototype supports two interaction modes. In \textit{automatic mode}, HDBSCAN is used to infer clusters from the document representation and to separate low-confidence points as an uncategorized group. In \textit{user-controlled mode}, Agglomerative Clustering assigns papers to a specified number of clusters, allowing users to inspect alternative granularities. Agglomerative Clustering was part of the offline evaluation (Section~\ref{sec:experiments}), where K-Means and Fuzzy C-Means were also evaluated to identify the ideal configuration for \textit{user-controlled mode}. HDBSCAN was not included in this evaluation because \textit{automatic mode} is intended to operate without requiring users to specify the number of clusters, and HDBSCAN has been used in previous work for theme discovery~\cite{weng_identification_2022, mersha_semantic-driven_2024}.


Cluster keywords are extracted using Class-Based TF-IDF (c-TF-IDF), inspired by BERTopic~\cite{grootendorst_bertopic_2022}. For each cluster, c-TF-IDF treats documents within a cluster as a single concatenated document and ranks terms by their importance relative to the other clusters, each likewise also represented as a single document~\cite{grootendorst_bertopic_2022}. The experimental evaluation also considers c-TF-IDF cluster keyword extraction for topic coherence scoring.

The interface visualizes each clustered paper by year and cluster identifier, shows intrinsic clustering metrics when available, and provides cluster tabs that expose titles, authors, categories, abstracts, and arXiv links.
The choice of text processing and clustering methods was guided by the need for a responsive user experience in the deployed prototype, while the offline evaluation explores a broader range of configurations to identify optimal settings for different scientific domains and query types, as detailed in Section~\ref{sec:experiments}.

The \textbf{public interface and release} is implemented in Streamlit~\cite{streamlit} which allows the retrieval and clustering pipeline to be deployed as an interactive Python web application. The released prototype supports keyword or phrase search, arXiv category selection, date filtering, sorting by relevance or submission date, automatic clustering, user-selected cluster counts, paper-level inspection, and source links.
To support reproducibility and transparency, the source code for the deployed prototype and evaluation scripts has been made publicly available\footnote{\url{https://github.com/ai4society/ELIOT}}. This repository includes the Streamlit application, data processing scripts, and configuration files for the offline evaluation. The code is organized to allow others to replicate the deployment and evaluation processes, ensuring that the results can be independently verified and extended in future work.

\section{Automated Offline Experiment}
\label{sec:experiments}
This section addresses \textbf{RQ1} through an automated offline configuration evaluation. The goal is to identify a robust default for interactive literature exploration; it is distinct from the user-facing evaluation in Section~\ref{sec:user_eval}, which studies perceived interpretability, usefulness, and use contexts.

\begin{table}[tbp]
    \centering
    \small
    \renewcommand{\arraystretch}{1.15}
    \begin{tabular}{l p{3.3cm} r}
        \toprule
        \textbf{Dataset Area} & \textbf{Topic}                       & \textbf{\#Docs} \\
        \midrule
        Computer Science      & Trustworthy AI                       & 300             \\
        Physics               & Condensed Matter (Superconductivity) & 253             \\
        Electrical Eng.       & Medical Image Processing             & 300             \\
        Mathematics           & Fluid Dynamics                       & 300             \\
        Statistics            & Statistical Machine Learning         & 300             \\
        Quant. Biology        & Neural Dynamics                      & 169             \\
        Quant. Finance        & Portfolio Optimization               & 252             \\
        Economics             & Econometric Modeling                 & 194             \\
        \bottomrule
    \end{tabular}
    \caption{Summary of arXiv datasets used for offline configuration evaluation. Electrical Eng.\ is shorthand for Electrical Engineering and Systems Science, and Quant.\ is shorthand for Quantitative.}
    \label{tab:datasets}
\end{table}

\subsection{Evaluation Setup}
Because users may search across very different domains, we evaluate configurations on eight arXiv datasets spanning various fields (Table~\ref{tab:datasets}). Each dataset is collected using a domain-specific query and arXiv category filter, ordered by relevance applied through the same retrieval layer used by the prototype. We cap each dataset at 300 papers where enough results are available for insightful analysis.

The \textbf{configurations} for the experiment are organized across three different axes.
We evaluate three document representation strategies: MiniLM (all-MiniLM-L6-v2)~\cite{reimers_sentence-bert_2019}, SciBERT~\cite{beltagy_scibert_2019}, and TF-IDF~\cite{qaiser_text_2018}. The TF-IDF vectorizer uses minimum and maximum document-frequency thresholds of $3$ and $0.7$, at most $3000$ features, and $1\mbox{-}3$ word n-grams.
We combine these representations with two dimensionality reduction methods, UMAP~\cite{mcinnes_umap_2018} and SVD~\cite{svd_zhang}, and three clustering algorithms with controllable $K$: K-Means~\cite{mcqueen1967some}, Agglomerative Clustering~\cite{ward1963hierarchical}, and Fuzzy C-Means~\cite{fuzzy_c_means}. For dimensionality reduction, we test $n_{\text{components}} \in \{5, 10, 15\}$. For clustering, we test $K \in \{3, 4, \ldots, 12, 15\}$; the value $15$ is included as a higher-granularity stress test without exhaustively searching every larger value.

Evaluation \textbf{metrics} are further divied into intrinsic clustering metrics (3) and topic coherence metrics (2). Silhouette Score (SIL)~\cite{rousseeuw1987silhouettes} measures how close points are to their assigned cluster relative to other clusters, with higher values preferred. Calinski-Harabasz Index (CHI)~\cite{calinski1974dendrite} measures the ratio of between-cluster separation to within-cluster dispersion, with higher values preferred. Davies-Bouldin Index (DBI)~\cite{davies1979cluster} measures average cluster similarity based on dispersion and separation, with lower values preferred.
Because these intrinsic metrics evaluate cluster geometry rather than label interpretability, we also compute topic coherence metrics using OCTIS~\cite{terragni_octis_2021} over the c-TF-IDF keywords assigned to each cluster~\cite{roder2015exploring}. $C_V$ estimates whether the top keywords in a cluster tend to co-occur in related contexts, while $C_{\text{NPMI}}$ measures normalized pairwise co-occurrence among keywords; for both metrics, higher values indicate more coherent and potentially more interpretable cluster labels.

Given the number of configurations, we use \emph{rank aggregation} to select the best configuration for each dataset. For each metric, configurations are ranked independently, with DBI ranked in ascending order and all other metrics ranked in descending order. We sum the metric ranks to obtain an aggregate score, where a lower score indicates a stronger configuration. We then collect the top configuration from each dataset and analyze the frequency of its components: representation, dimensionality reduction method, number of components, and clustering algorithm. The final recommended configuration is determined by the most frequent components, with average metric performance used to resolve ties.

\begin{table*}[t]
    \centering
    \small
    \renewcommand{\arraystretch}{1.6}
    \begin{tabular}{l l l c c l c c c c c}
        \toprule
        \textbf{Dataset}     & \textbf{Representation} & \textbf{Dim. Red.} & \textbf{$n_{comp}$} & \textbf{K} & \textbf{Algorithm} & \textbf{CHI} & \textbf{DBI} & \textbf{SIL} & \textbf{$C_V$} & \textbf{$C_{\text{NPMI}}$} \\
        \midrule
        Computer Science     & MiniLM             & UMAP               & 10                  & 3          & K-Means            & 310.29       & 1.02         & 0.58         & 0.47           & -0.27                      \\
        Physics              & MiniLM             & UMAP               & 15                  & 10         & K-Means            & 129.48       & 1.07         & 0.51         & 0.45           & -0.26                      \\
        Mathematics          & MiniLM             & UMAP               & 5                   & 4          & Fuzzy C-Means      & 406.23       & 0.91         & 0.57         & 0.73           & -0.21                      \\
        EESS                 & MiniLM             & UMAP               & 15                  & 9          & Agglomerative      & 145.65       & 0.93         & 0.48         & 0.47           & -0.21                      \\
        Statistics           & MiniLM             & UMAP               & 10                  & 6          & Agglomerative      & 105.21       & 0.98         & 0.47         & 0.38           & -0.28                      \\
        Quantitative Biology & MiniLM             & UMAP               & 10                  & 10         & Agglomerative      & 65.97        & 0.98         & 0.47         & 0.43           & -0.21                      \\
        Quantitative Finance & MiniLM             & UMAP               & 10                  & 4          & Agglomerative      & 528.42       & 0.80         & 0.59         & 0.58           & -0.18                      \\
        Economics            & MiniLM             & UMAP               & 10                  & 11         & K-Means            & 123.94       & 0.86         & 0.59         & 0.38           & -0.28                      \\
        \bottomrule
    \end{tabular}
    \caption{Top-ranked configuration of each dataset selected via rank aggregation over clustering and coherence metrics. EESS is shorthand for Electrical Engineering and Systems Science.}
    \label{tab:best_configs}
\end{table*}

Note that this evaluation is designed to identify robust defaults, not to prove that one clustering method is universally optimal. The datasets are query-dependent, arXiv-centered, and limited to titles and abstracts. The coherence metrics evaluate keyword interpretability indirectly and should be complemented by user judgments. These limitations motivate the user study and focus group described in Section~\ref{sec:user_eval}.

\subsection{Evaluation Results}
\label{sec:results}

The experiment results answer \textbf{RQ1}. Table~\ref{tab:best_configs} summarizes the top-ranked configurations for each dataset, selected via rank aggregation over clustering and coherence metrics, explained in Section~\ref{sec:experiments}.
The coherence scores should be interpreted comparatively across configurations rather than as absolute quality judgments, since they are computed from short title-abstract texts and automatically extracted cluster keywords.

Regarding \textbf{text representation}, MiniLM appears in the top-ranked configurations across all eight datasets, while neither SciBERT nor TF-IDF appears as the best-performing representation.
This result is promising given the task. MiniLM is trained to produce sentence-level embeddings that capture semantic similarity, making it well-suited for clustering title-abstract representations.
This result also clarifies how \systemname{} differs from the \emph{LLM-planning} script~\cite{pallagani2025revisiting}. In that setting, TF-IDF was used for supervised classification into a small, expert-defined taxonomy, where category labels and training examples already constrained the task. In \systemname{}, cluster boundaries are not predefined and terminology varies across domains, so lexical overlap is a weaker signal than semantic similarity.
In contrast, SciBERT produces contextualized token-level representations and often benefits from task-specific pooling or fine-tuning for downstream use. TF-IDF is fast and transparent, but it relies on surface word overlap, which can miss semantically related papers that use different terminology~\cite{weng_identification_2022}.

About \textbf{dimensionality reduction and clustering}, UMAP is present in all top-ranked configurations, while SVD does not appear in any of them. This suggests that nonlinear dimensionality reduction better preserves the structure useful for clustering in these datasets. Among UMAP configurations, $n_{\text{components}}=10$ is the most frequent value.
Agglomerative Clustering appears most frequently among the top configurations, followed closely by K-Means. The difference is modest, so the result should be interpreted as a default-selection signal rather than a decisive algorithmic result. For an applied interactive system, this still matters: a stable default reduces the need for users to understand clustering hyperparameters before beginning exploration.

Based on these findings, the default \systemname{} configuration uses MiniLM, 10-dimensional UMAP, and Agglomerative Clustering, since each component occurs most frequently among the top-ranked configurations across datasets. This answers \textbf{RQ1} by selecting a practical default from the tested configuration space, rather than claiming universal optimality.

The strongest applied takeaway is that \systemname{} is usable with its 
 defaults selected from a benchmark-driven discovery process, even though it uses off-the-shelf methods.
We do not assume that users will tune representation models, dimensionality reduction, and clustering algorithms. 
The system should therefore expose simple default interaction choices while internally relying on a configuration that generalizes across several domains. These automated results support the machine-learning component of \systemname{} as a practical structuring layer for retrieved literature, but they do not by themselves validate whether users find the clusters interpretable or useful; those questions are addressed through the user evaluation in the following section.

\section{User Evaluation}
\label{sec:user_eval}

This section addresses \textbf{RQ2} and \textbf{RQ3} through two complementary methods: a scenario-based survey study and an expert focus group. Unlike the automated offline evaluation in Section~\ref{sec:experiments}, this evaluation studies user-facing interpretability, perceived usefulness, and the contexts in which researchers expect \systemname{} to be most valuable.

We designed the \textbf{survey-based user study} to evaluate usefulness, interpretability, and interaction flow for \textbf{RQ2}. Participants were asked to evaluate \systemname{} using automatic clustering mode for two predefined scenarios. In \textbf{Scenario 1}, they reviewed a Computer Science search with the query \emph{large language models} over the April $2024\mbox{-}2026$ period. In \textbf{Scenario 2}, they reviewed a Quantitative Finance search with the query terms \emph{market} and \emph{prediction} over the same period.
Each scenario presented retrieved papers, cluster keywords, and temporal views. Participants then used \systemname{} with their own queries in a self-directed task. Both the predefined scenarios and open-ended interaction were evaluated with 5-point Likert-scale questions and free-response prompts.
The complete survey instrument, including the exact scenario configurations, representative cluster keywords, visualizations, survey questions, and free-response prompt, is available as supplementary material.\footnote{\url{https://drive.google.com/file/d/1NA3mcGRJ4obaFIudOhanZGfbm1Y1vNnV}}


A total of 10 respondents, consisting of graduate students and faculty members at a public university, participated in the survey.
Across the two predefined scenarios, participants perceived the group labels as meaningful; 85\% rated 4 or 5. A one-tailed Wilcoxon signed-rank test rejected the null hypothesis that users are neutral regarding label meaningfulness ($p < 0.001$). Since the respondents may not share their backgrounds with the example scenarios, the perceived usefulness of the clusters was moderate, with 50\% of the responses rated 4 or 5. The difference from neutral was not statistically significant according to the same test ($p = 0.137$). To examine this pattern further, we convened a {\em focus group of experienced researchers} whose feedback we discuss next. In addition, participants showed a positive perception that \systemname{} could be useful for their future research ($p = 0.047$).


To study \textbf{RQ3}, we convened a \textbf{focus-group-based user evaluation} consisting of four experienced researchers at a public university. They all held PhDs, and together had research experience of over six decades in computer science and engineering (with two having experience of less than 10 years each and two more than 25 years each). In a recorded session, we reviewed their experience with the literature exploration problem in recent years, explained \systemname{}, demonstrated its capabilities, and asked for feedback.

Overall, the participants were very positive about the deployed system and how they may use it for their research. They liked the overall design and performance of the tool, the ability to interact with cluster visualizations, get information about cluster quality and labels (descriptions), and directly access papers. They also suggested areas of improvement, including making the details and limitations of the data source explicit (e.g., arXiv's API limits and category-filter recall), and providing information about query capabilities. In particular, one concern was that a query restricted to one arXiv category can miss relevant papers assigned to another category or distributed across multiple categories. This is a data-source constraint faced by arXiv-based tools generally, rather than a limitation specific to \systemname{}'s clustering pipeline, and thus, classified as a future improvement to the tool's documentation and user interface rather than a change to the underlying system. 

The studies reveal important \textbf{findings}. With respect to \textbf{RQ2}, the survey results suggest that users find the generated labels meaningful and see promise for future research use, while perceived usefulness of predefined clusters is more dependent on the user's background and the scenario topic. With respect to \textbf{RQ3}, the focus group suggests that \systemname{} is especially useful when researchers need a traceable overview of fast-changing technical areas. This evidence is preliminary and qualitative rather than a controlled comparison between established and emerging research areas.

These findings connect back to the \emph{LLM-planning} studies that motivated \systemname{}~\cite{pallagani2024prospects,pallagani2025revisiting}. Those studies required expert taxonomies and manual or semi-automated updates to make category shifts interpretable. The user feedback suggests that \systemname{} generalizes the same need: meaningful labels, cluster inspection, quality information, and direct access to source papers help make evolving literature categories interpretable and auditable even when users begin from a query rather than a predefined taxonomy.

We also noticed that users were sometimes confused when the same query produced substantially different cluster granularities in {\em automatic mode} across different time windows, such as one year versus three years. This can occur because \systemname{} caps retrieved documents by default (300), and wider time windows with a fixed fetch limit may change the density structure in the embedding space used by HDBSCAN in {\em automatic mode}. To address this limitation, \systemname{} exposes two complementary controls. First, users can increase the maximum number of retrieved papers; in our observation, increasing this limit mitigated the issue. Second, users can switch to {\em user-controlled mode} and specify the desired number of clusters directly, with the system's suggested value as an optional starting point.


\section{Conclusion}

We presented \systemname, a public prototype for exploring evolving scientific literature through online retrieval, clustering, keyword labeling, temporal visualization, and document-level inspection. The system is motivated by a practical need first encountered in our \emph{LLM-planning} literature studies: researchers working in fast-moving areas need traceable overviews of changing corpora, not only ranked lists, opaque generated summaries, or one-off scripts built around a fixed taxonomy.

Our evaluation separated automated configuration selection from user-facing assessment. For \textbf{RQ1}, MiniLM and UMAP appeared in every top-ranked configuration, and Agglomerative Clustering appeared most often, motivating MiniLM with 10-dimensional UMAP and Agglomerative Clustering as the default. For \textbf{RQ2}, the survey results indicate that users generally found the cluster labels meaningful and saw promise for future research use, while cluster usefulness depended more on scenario fit. For \textbf{RQ3}, the expert focus group suggested that \systemname{} is most valuable as a traceable overview tool for fast-changing technical areas, while also surfacing limitations around data-source transparency, query guidance, and interpreting clusters across different time windows.

Several direction remain for future.
On the retrieval side, query expansion via WordNet-like lexical resources~\cite{miller1995wordnet}, domain ontologies~\cite{bhogal2007review}, relevance feedback, or LLM-suggested terms could broaden recall before clustering, while keeping the explicit-query design that supports traceability.
Multi-source retrieval from additional scholarly platforms, such as medRxiv~\cite{medrxiv}, ResearchGate~\cite{researchgate}, or broader scholarly indexes where available, could also reduce dependence on arXiv-specific category boundaries.
On the evaluation side, the offline configuration study is limited to titles and abstracts from eight arXiv domains; extending it to full texts or other data sources would test the robustness of the recommended defaults.
Further, while Streamlit and its free hosting tier served the prototype well, custom hosting and larger web-development frameworks could improve \systemname{}'s customizability and avoid scale-to-zero, which forces the app to sleep after a few hours.
Finally, the user study and focus group reported here are preliminary; a larger-scale evaluation with controlled tasks would provide stronger evidence for this practical, auditable workflow for turning query-time retrieval into structured literature views that researchers can inspect, question, and refine.

\section{Acknowledgments}
We thank Aarohi Goel and Thrinadh Sai for contributions to an earlier version of \systemname, and participants in our user studies. 


\section*{GenAI Usage Disclosure}
The authors used generative AI tools for limited language editing, organization, and phrasing assistance during manuscript preparation. Generative AI tools were also used to support prototype development and to generate the illustrative cluster image shown in step 3 of Figure~\ref{fig:system_pipeline}. All AI-assisted text, image, and code were reviewed, edited, and verified by the authors. The authors take full responsibility for the paper's content, implementation, and conclusions. Generative AI was not used to collect data, run experiments, compute evaluation results, or draw scientific conclusions.


\printbibliography

\end{document}